\documentclass[9pt,onecolumn,twoside]{article}
\usepackage[utf8]{inputenc}

\usepackage{amsmath}

\usepackage{graphicx}

\usepackage{amssymb}
\usepackage{authblk}

\author[*,1]{Stefan Rothe}
\author[1]{Nektarios Koukourakis}
\author[1]{Hannes Radner}
\author[2]{Andrew Lonnstrom}
\author[2]{Eduard Jorswieck}
\author[1]{Jürgen W. Czarske}

\affil[1]{Technische Universität Dresden, Faculty of Electrical and Computer Engineering, Laboratory of Measurement and Sensor System Technique, 01062 Dresden, Germany}
\affil[2]{Technische Universität Braunschweig, Faculty of Electrical Engineering, Information Technology, Physics, Institute for Communications Technology, 38106 Braunschweig, Germany}

\affil[*]{Corresponding author: stefan.rothe@tu-dresden.de}
\date{}                     
\title{Physical Layer Security in Multimode Fiber Optical Networks}

\begin{document}

\maketitle

\section{Abstract}
Inverse precoding algorithms in multimode fiber based communication networks are used to exploit mode dependent losses on the physical layer. This provides an asymmetry between legitimate (Bob) and unlegitimate (Eve) receiver of messages resulting in a significant SNR advantage for Bob. In combination with dynamic mode channel changes, Eve has no chance to reconstruct a sent message even in a worst case scenario in which she is almighty. This is the first time, Physical Layer Security in a fiber optical network is investigated on the basis of measured transmission matrices. These results show that messages can be sent securely  with conventional communication techniques. Translating the task of securing data from software to hardware represents the potential of a scientific paradigm shift. The introduced technique is a step towards the development of cyber physical systems.

\section{Introduction}

\noindent The number of exchanged data via internet has increased exponentially in recent years \cite{cisco2015}. Following this trend, the amount of sensitive data has increased in the same way and thus the importance of information security. Today the most commonly used technique warranting secure communication is based on digital cryptographic keys, which exploit the complexity of multiplying large prime numbers \cite{amitonova2018}.  However, cryptographic algorithms are facing several challenges. First, they are vulnerable against unexpected technological developments, as parallel networks of computers are breaking codes that have been considered safe in polynomial time using high-performance computers \cite{pappu2002,shor1997}. Second, the demands which are addressing the physical embodiments of these functions go beyond the constraints of conventional semiconductor technology. Third, a thief stealing a digital key can go unnoticed.

Overcoming the drawbacks of computational cryptography, investigations of physical cryptographic methods have been made \cite{Ruhrmair2012,Javidi2016,Situ2007,Javidi2000}. Physical unclonable functions (PUFs) are physical objects, that can not feasibly be copied due to their comprehensive number of degrees of freedom. Therefore, they have been studied as a physical one-time pad \cite{horstmeyer2013} and as a secure physical authentication protocol \cite{goorden2014,uppu2018,pappu2002,Mesaritakis2018}, respectively. In both scenarios optical PUFs were generated by illuminating scattering media. 

One-time pads are information theoretically secure, however the realization is highly impractical as both the sender (e.g. \textit{Alice}) as well as the receiver (e.g. \textit{Bob}) of a message cannot synchronize their channel without having exactly the same \textit{unique} PUF.

In applications where PUFs were utilized to create secure authentication, the security relies on the difficulty of cloning the optical response \cite{pappu2002}, as well as in combination with a low mean photon number \cite{goorden2014,uppu2018}. This physical encryption technique always assumes that Alice and Bob have access to the PUF in an initial calibration step, which is not observed by a possible eavesdropper (e.g. \textit{Eve}). This is a major problem, as due to statistical fluctuations in the light path caused by temperature, mechanical stress and local phase-shifts, the PUF is changing in time. Consequently, Alice and Bob need to recalibrate their channel which is not practical when Eve is not supposed to watch the calibration procedure. 

Quantum Key Distribution (QKD) is another technique which addresses the secure transmission of data. QKD utilizes the inimitability of unknown quantum states to make the reconstruction of encrypted messages impossible \cite{bennett2014,Frohlich2017}. Even though QKD offers an unconditionally secure data transmission, serious problems arise when combining QKD signals with optical amplifier noise and classical communications \cite{winzer2015,peters2010} on a conventional fiber optical infrastructure.

The use of multimode fibers (MMF) in fiber optical networks is regarded as a promising approach to increase possible data rates significantly. However, the phenomenon of mode scrambling inside an MMF was considered as a hurdle for the MMF usage for a long time. Once coherent light is sent into the MMF on one side of the fiber, it will appear as a granulated structure on the other side of the MMF called speckle pattern. This barrier was overcome with the development of 
wavefront shaping (WS) \cite{Vellekoop2007}. 
Firstly, it became feasible in optical engineering to control the propagation of light through fluctuating surfaces by WS \cite{Koukourakis2016}, 
secondly the light control through MMFs has not only become possible, but much more important. Nowadays, the MMF is a key device in several fields of research. They are used in biophotonical applications \cite{Papadopoulos2012,Czarske2016,Haufe2017,Gu2015,Ploschner2015,Loterie2015,Leite2018,Turtaev2018} to gain access to hard-to-reach areas due to their flexibility and high number of degrees of freedom in a minimum space. These properties are particularly helpful for image transmission using the MMF as an ultrathin endoscope. In addition, achievable data rates in communication networks can be significantly increased using MMFs, since they can be used to develop novel multiplexing techniques in which space is a scalable parameter \cite{Berdague1982,Richardson2013,Ryf2015,Carpenter2016}.

Further, the property of light scrambling inside an MMF has opened the door to increase the level of information security as the scattering characteristics are completely random. Therefore, by controlling the transmission channel between Alice and Bob using WS, Eve, who is tapping off the signal somewhere in the center of the fiber, will only receive a scrambled speckle pattern. Additionally, this idea has been investigated in combination with a low photon number, so that Eve only receives a fraction of information \cite{amitonova2018}. However, The approach presented has two major drawbacks. On the one hand, a low-photon source is used, which is impractical with regard to desired transmission distances in fiber optical networks. On the other hand, it is assumed that Eve has no access to the transmission channel between Alice and Bob during the calibration phase. For realistic scenarios it is unknown whether an eavesdropper is able to witness the calibration or not. Thus a data transmission can only be considered secure, when Eve is allowed to watch the calibration.

In this paper, a novel approach to enhance the information security in fiber optical networks using physical layer security (PLS) is introduced. In this approach a step back to the basic understanding of the underlying channel model is made \cite{Jorswieck2010}. As introduced in \cite{guan2015}, the modes supported by the MMF are unevenly exiting the MMF if someone is tapping off light between Alice and Bob. This phenomenon arises from the physics of the MMF \cite{ho2011} and is called mode dependent loss (MDL), which is the key assumption of the introduced model. Consequently, in the developed setup a conventional coherent light source is used. The attacker Eve is allowed to have access to the communication channel during the calibration phase between Alice and Bob. Thus, Eve has knowledge of the transmission matrix ($T_M$) between Alice and her, as well as the $T_M$ between Alice and Bob. Eve's supremacy, which is assumed in the introduced method, represents the main difference between PLS and other techniques designed to ensure information security.

\section{Measurement of the Multimode Fiber's Transmission Matrix}
The $T_M$ approach proved to be extremely helpful to overcome the challenge to use the MMF either as an imaging \cite{Gu2015,Ploschner2015,Loterie2015,Leite2018,Turtaev2018} or data transmitting tool \cite{Carpenter2014,Carpenter2016} for compensating the fiber's light scrambling property. Besides experimental techniques to determine the $T_M$, approaches to identify the $T_M$ via deep learning \cite{Rahmani2018,Borhani2018} or convex optimization algorithms \cite{NGom2018} have already been shown.

The $T_M$ describes the linear and complex relationship between input and output of the MMF exactly. Hence it is possible to compensate the mode scrambling process during light transmission by inverting the $T_M$ to pre-distort the incident light signal \cite{Carpenter2014, Carpenter2016, Loterie2015}. Consequently, the $T_M$ offers excellent possibilities to control the light transmission through the MMF. Furthermore, knowledge of the $T_M$ of the MMF is required for performance characterization employing Mode Division Multiplexing (MDM) in order to recover channels by optical means \cite{Carpenter2014}. 

There are several ways possible to acquire a $T_M$ of an MMF experimentally. In a first step, a set of linear independent patterns, which sufficiently describe light entering and exiting the MMF, need to be chosen. These patterns are the basis of the $T_M$. Generally, a WS element like a spatial light modulator (SLM) is used to spatially shape an incident Gaussian laser beam imprinting each pattern of the set onto it. Afterwards, the modulated light field is imaged onto the input fiber facet where a certain group of modes of the MMF's mode domain is excited and propagates through the MMF. The light, which is exiting the MMF has to be measured and decomposed into the domain of the previously chosen set of independent patterns. After this process has been repeated for every pattern of the set, the measured complex decomposition coefficients are stored row by row forming the $T_M$. 

In 2010 the results of the first $T_M$ measurement in optics have been published \cite{Popoff2010}, where the light propagation through a random scattering sample was investigated. By using a phase-only SLM, Hadamard patterns, whose elements are either $+1$ or $-1$ in amplitude, were chosen as the basis for the $T_M$. However, this procedure was developed for a sample having generalized scattering properties. There are more practical ways to acquire the $T_M$ of a MMF depending on the targeted application. Using the MMF as an flexible endoscope, the aim is to generate scanning focal spots at the end of the fiber. Therefore it is obvious to 
choose diffraction limited focal spots \cite{Ploschner2015,Turtaev2018,Leite2018} as the $T_M$ basis. Another suitable basis to describe light transmission through an MMF is a basis of plane waves with varying spatial frequencies \cite{Loterie2015}. One major advantage in this approach is the simplicity of the Fourier transform relationship between the plane wave basis and the physical pixel basis of the SLM, respectively the camera. However, 
in terms of measurement and computational effort, it is beneficial to work with the smallest possible basis. In MMF applications the smallest possible basis is represented by the MMF's mode domain with the number of $N$ modes. Both the plane wave and the focal spot as well as the Hadamard bases are larger than $N$. In \cite{carpenter2012,Carpenter2014,Carpenter2016}, the chosen bases are the MMF's mode domain, respectively. They use two SLMs, one on each side of the MMF. One SLM is used to adaptively generate complex light fields spatially according to each mode of the MMF's mode domain for exciting the modes the MMF supports sequentially. The modulated light fields are shaped in amplitude and phase with a phase-only SLM by applying special phase masks using the simulated annealing method \cite{Kirkpatrick1983}. The second SLM is then used to demultiplex the scrambled light field based on a correlation filter method, where the efficiency of diffracted light off the SLM surface into a power meter is proportional to the correlation of the fiber output with a desired mode. The given measurement effort per decomposition process is $4N-3$ per polarization \cite{carpenter2012}. 

\begin{figure}[htb]
    \centering
    \includegraphics[width=0.6\textwidth]{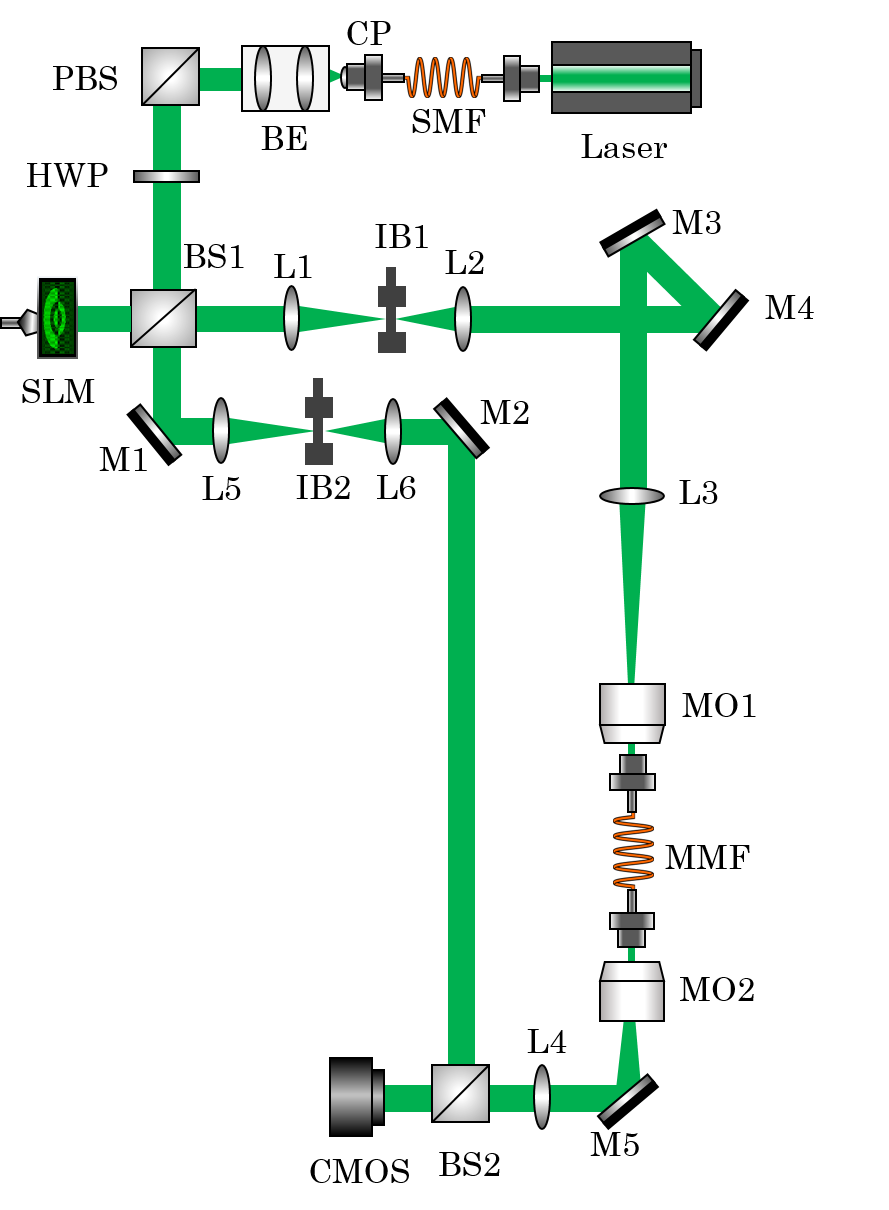}
    \caption{Optical setup, which is used to measure the $T_M$ of an MMF.}
    \label{fig:setup}
\end{figure}

In this paper, the optical setup [Fig. \ref{fig:setup}] which has been introduced in \cite{Rothe2019} is used to measure the $T_M$ of a step-index MMF (THORLABS M68L, $\varnothing25\mu$m, NA=0.1) with exactly $N$ measurements. The utilized light source is an 532nm solid state laser (LaserQuantum, TORUS). This combination leads to a mode domain of 55 modes per polarization direction. Mode selective excitation is performed using amplitude and phase modulation based on superpixel phase masks with a single spatial light modulator (HOLOEYE Pluto, 8bit phase-only SLM). The light that propagates through the MMF is holographically measured using conventional off-axis digital holography. The measured complex light field is then decomposed into the MMF's mode domain. For this step, the orthogonality of the mode field distributions in the mode domain is exploited and a complex series expansion is carried out. Using this technique, one row of the $N\times N$ $T_M$ is measured in single-shot. 
\section{Recovering Individual Transmission Channels of Multimode Fibers Using Transmission Matrix Inverse Precoding}
\begin{figure}[htb]
\centering
\includegraphics[width=0.6\textwidth]{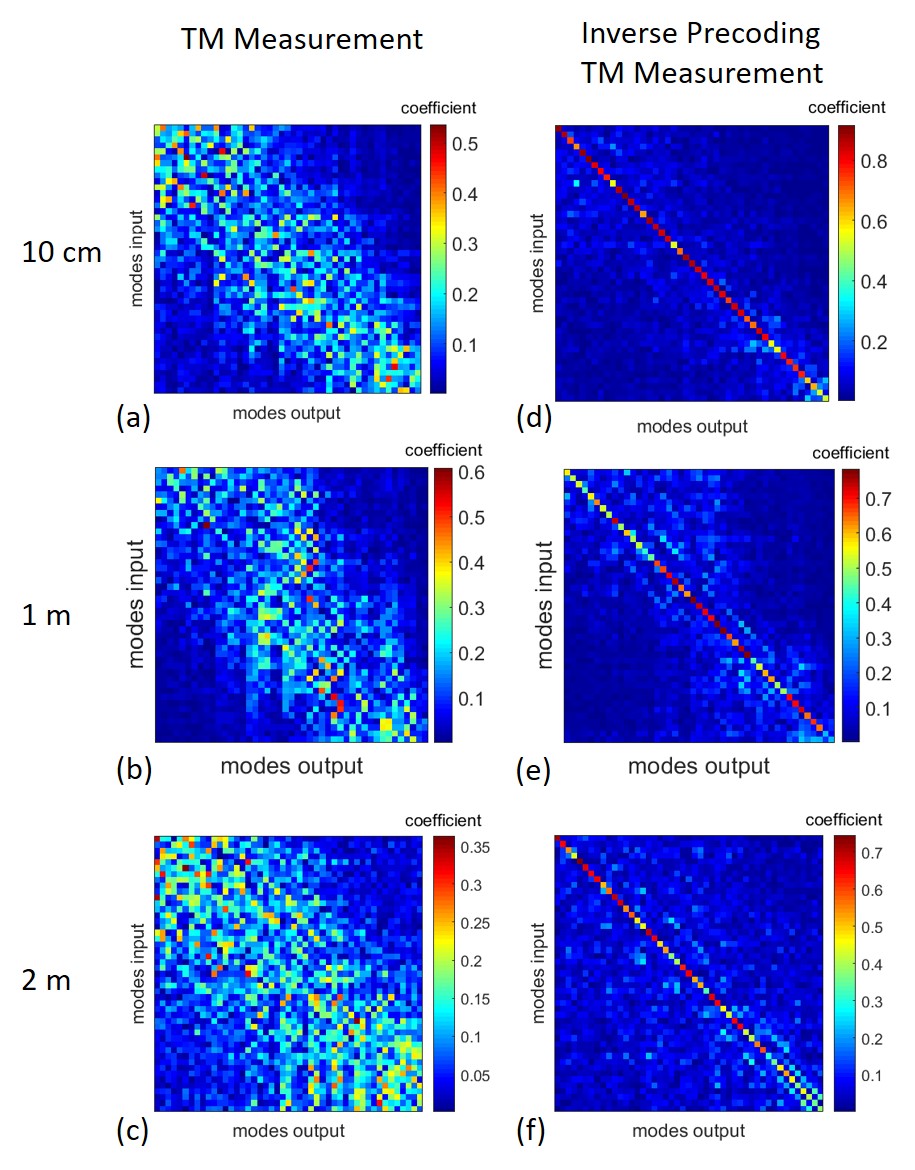}



\caption{(a)-(c) $T_M$ measurements of step-index MMF's of different lengths, as well as (d)-(f) pre-coded diagonalized $T_M$ measurements. The individual images show the pure amplitude values of the complex-valued $T_M$s. The $LP_{l,m}$ modes are sorted ascending first by their $l$ index and then by their $m$ index: $LP_{01}, LP_{02}, \cdots LP_{11}, \cdots LP_{LM}$. (a)-(c) $T_M$ measurements of 10~cm, 1~m and 2~m MMFs, respectively. (d)-(f) diagonalized $T_M$s of 10~cm, 1~m and 2~m MMFs, respectively.}
\label{fig:$T_M$}
\end{figure}

Once the $T_M$ of an MMF is measured, the complete light transmission characteristic between input $\vec{x}_{A}$ on Alice's side and output $\vec{y}_{B}$ on Bob's side of the MMF is determined:
\begin{equation}
    \vec{y}_{B}=T_{M,AB} \cdot \vec{x}_{A} + n_{B},
    \label{eq:transmission}
\end{equation}
where $T_{M,AB}$ is the $T_M$ between Alice and Bob and $n_B$ is the measurement noise on Bob's side. The input vector $\vec{x}_{A}$ is defined as follows to ensure an average constant transmission power:
\begin{equation}
    \vec{x}_{A}=\frac{\vec{x}_{A}}{||\vec{x}_{A}||}.
\end{equation}
Alice is able to undo the scrambling property of the MMF by inverting the $T_M$. Tikhonov inversion is used for inverting the $T_M$, because it is robust against influence of noise. In the following, inverted matrices are marked with a superscript $\dagger$. The Tikhonov inversion is performing an regularization process and has already been used in biophotonic applications \cite{Loterie2015,Popoff2010}. The regularization parameter has been chosen as $12\%$ of the maximum singular value of the $T_M$. After superimposing the inverted $T_M$ $T^{\dagger}_{M,AB}$ to the input signal $\vec{x}_{A}$, the new input $\hat{\vec{x}}$ can be described as:
\begin{equation}
    \hat{\vec{x}}_{A}=T_{M,AB}^{\dagger}  \vec{x}_{A},
    \label{eq:transmissionNew}
\end{equation}
with 
\begin{equation}
\hat{\vec{x}}_{A}=\frac{\hat{\vec{x}}_{A}}{\sqrt{tr\{T_{M,AB}^{\dagger}T_{M,AB}^{\dagger,H}\}}}.
\end{equation}
The superscripted $H$ indicates a Hermitian transpose operation. Replacing $\vec{x}_{A}$ of equation (\ref{eq:transmission}) with $\hat{\vec{x}}_{A}$ of equation (\ref{eq:transmissionNew}), it is possible for Bob to observe the signal directly without performing any signal processing according to the model equations
:
\begin{equation}
    \begin{split}
        \vec{y}_{B}& =T_{M,AB}  \hat{\vec{x}}_{A} + n_{B} \\
                   & =T_{M,AB}  T_{M,AB}^{\dagger}  \vec{x}_{A} + n_{B} \\
                   & =\vec{x}_{A}+n_{B}. \\
    \end{split}
    \label{eq:bob}
\end{equation}
Using the new input signal calculation rule [Eq. (\ref{eq:transmissionNew})], Alice gets the required combinations of complex mode weights to make a specific output signal appear on Bob's side. Shaping such complex light field distributions requires an adaptive optical device with a high modulation depth like an SLM. If the $T_M$ is measured again using the mode domain base with inverse precoding, the amplitude distribution of the $T_M$ will be very similar to an identity matrix. In Figure \ref{fig:$T_M$} $T_M$ measurements of MMFs of different lengths are shown. The efficiency of the inverse precoding process $\eta_{p}$ can be quantified by calculating the mean optical power, which is located on the diagonal elements of the diagonalized $T_M$s $\overline{P}_{d}$ with respect to the mean optical power, which is distributed over the background entries $\overline{P}_{b}$ \cite{Ploschner2015}:
\begin{equation}
\eta_{p}=\frac{\overline{P}_{d}}{\overline{P}_{d}+\overline{P}_{b}}
\end{equation}
The efficiencies of the individual precoding processes shown in Figure \ref{fig:$T_M$} are distributed as follows: the precoding of the 10cm MMF in Figure \ref{fig:$T_M$}(d) has an efficiency of $95\%$, while the efficiency of the 1m MMF Figure \ref{fig:$T_M$}(e) was $90\%$ and $89\%$, respectively, for the 2m MMF in Figure \ref{fig:$T_M$}(f). These results show that the inverse precoding process can also be efficiently performed on MMFs up to 2m in length. Thus, the mean transported power in the individual channels can be transmitted with an efficiency of at least $89\%$.
\section{Parallel Mode Division Multiplexing of multiple individual spatial fiber modes}
\begin{figure}[htb]
\centering
\includegraphics[width=0.7\textwidth]{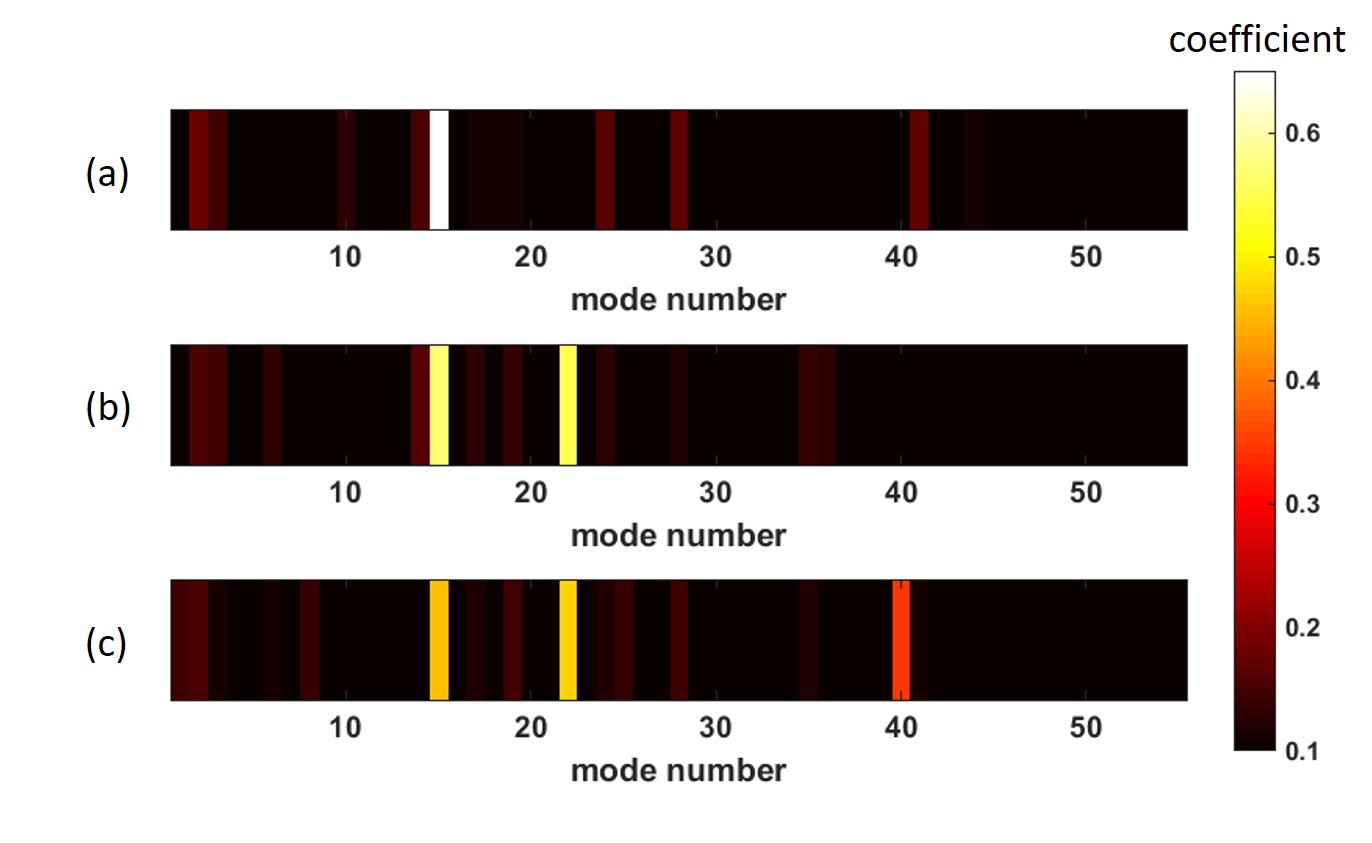}
\caption{Parallel MDM of (a) one, (b) two and (c) three individual spatial fiber modes of an 1m step-index MMF.}
\label{fig:MDM}
\end{figure}

The diagonalization of the MMF by means of inverse precoding not only has the advantage that Bob receives Alice's message directly without having to perform further signal processing, but since the individual spatial fiber modes were selected as the basis of the $T_M$, MDM can be employed. In addition, by 
superimposing superpixel phasemasks which are generating individual complex mode distributions on Bob's side, Alice is able to 
transmit Data on multiple channels simultaneously as shown in Figure \ref{fig:MDM}. By the transmission of a light signal of constant power over several channels, the power is distributed to the individual channels. As a result, the average signal level decreases with increasing number of channels. The magnitude of the received mode coefficient in Figure \ref{fig:MDM}(a) is approximately 0.62, while the average magnitude is approximately 0.4 in Figure \ref{fig:MDM}(c). This constitutes a performance limitation of MDM in this system, which could be compensated by increasing the laser power. Nevertheless, assuming that simple thresholding is used, Alice can now choose between $55\times54\times53=157.410$ transferable symbols. This corresponds to a 17 bit transmission system. Using the SLM with an repetition rate of 60~Hz, Alice could achieve a transmission rate of approximately 7.5~Mbit with a cw laser of only one wavelength.
\section{Realization of Physical Layer Security in Fiber Optical Networks}
\begin{figure}[htb]
\centering
\includegraphics[width=0.7\textwidth]{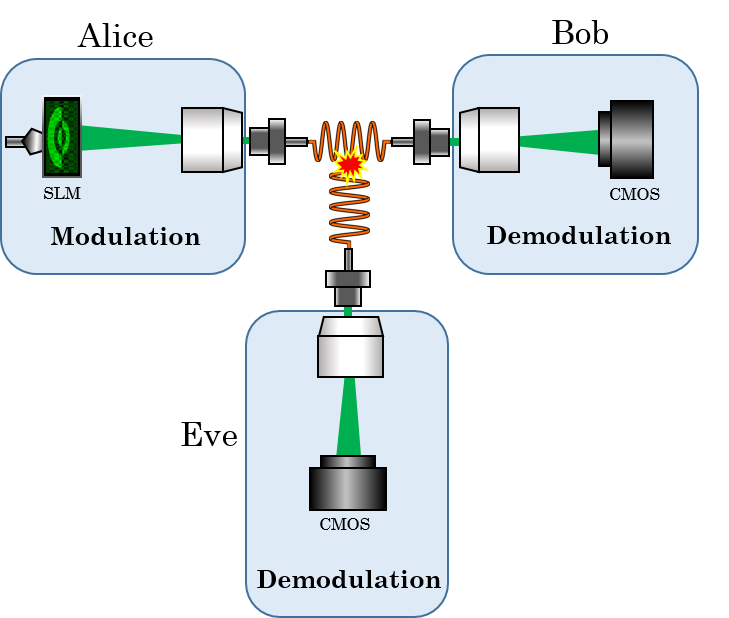}
\caption{Eavesdropping scenario. Eve splices herself on the MMF used as a transmission channel between Alice and Bob. She will already listen during the calibration phase between Alice and Bob and thus knows the $T_M$ between Alice and Bob. The light which is exiting on Eve's side is of sufficient intensity, but so low that Eve's listening process is not registered. She is able to examine the light field distribution e.g. holographically.}
\label{fig:attack}
\end{figure}

As already mentioned in detail in the introduction, in the communication scenario presented here Eve is given access to the transmission channel between Alice and Bob from the beginning. Therefore, she has the possibility to measure both the $T_M$ between Alice and Bob and the $T_M$ between Alice and her. She gains access to the transmission channel by tapping into the fiber via a splicing process without interrupting the signal as shown in Figure \ref{fig:attack}. Like Bob, she will also evaluate the light signal emitted from the MMF holographically. 

\noindent The light transmission process between Alice and Eve can be described as follows:
\begin{equation}
    \begin{split}
        \vec{y}_{E}& =T_{M,AE}  \hat{\vec{x}}_{A}+n_{E}\\
        & =T_{M,AE} T_{M,AB}^{\dagger} \vec{x}_{A} +n_{E}
    \end{split}
    \label{eq:transmissionEve}
\end{equation}
whereas the noise level on Bob's side $n_{B}$ has the same level as on Eve's side $n_{E}$. According to \cite{winzer2015} the probability for coupling modes into Eves tapping fiber is mode dependent. This process can be described using a diagonal matrix $V$, which is inserted into the transfer function, which carries the mode-dependent power loss coefficients $\sigma_i^2$ on the diagonal elements representing the MDL characteristic of the MMF:
\begin{align}
V = 
\left[
  \begin{matrix}
  \sigma_1^2 &   0  & 0   \\
  0 & \ddots & 0     \\
  0 & 0 & \sigma_N^2   \\
	\end{matrix}
\right],
\end{align}
where $\sigma_{min}<<1$ and $\sigma_{max}=1$, which results in the following transmission relation:
\begin{equation}
    \vec{y}_{E} =\sqrt{V} T_{M,AE} T_{M,AB}^{\dagger} \vec{x}_{A} +n_{E}
\end{equation}
As soon as Eve wants to decode an observed message $\vec{y}_{E}$, she has to invert the measured $T_M$ $H= \sqrt{V}T_{M,AE} T_{M,AB}^{\dagger}$:
\begin{equation}
    \begin{split}
        \Tilde{y}_{E} & =H^{\dagger} \vec{y}_{E}\\
        & =\vec{x}_{A}+H^{\dagger}n_{E},
    \end{split}
    \label{eq:outputeve}
\end{equation}
with 
The introduced relationships lead to two important observations:
\begin{enumerate}
    \item Due to the fact that values between 0 and 1 are located on the diagonal elements of $V$, the entries of $H$ are attenuated. This will lead to noise amplification during Eve's inversion process. 
    \item Due to Alices's inverse precoding, Alice directly influences Eve's noise term [Eq.(\ref{eq:outputeve})
    ], but not Bob's noise term [Eq. (\ref{eq:bob})].
\end{enumerate}
Possible influences of the two presented findings are examined in more detail in the following simulation model.
\section{Security Analysis of a Multimode Fiber Optical Network using Physical Layer Security}
\begin{figure}[htb]
    \centering
    \includegraphics[width=0.7\textwidth]{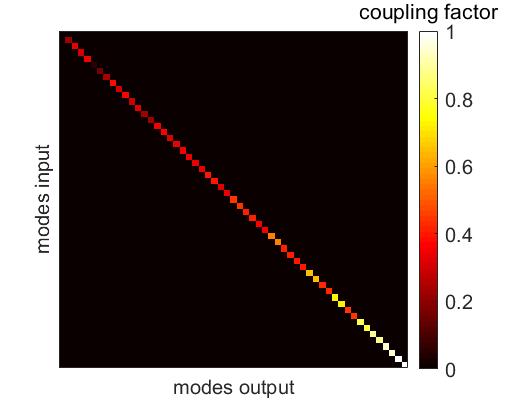}
    \caption{$V$ matrix representing the coupling characteristic for Eve. The trend of the diagonal entries was chosen proportional to the power ratio in the edge area to the total power of the core of the respective modes.}
    \label{fig:V}
\end{figure}

\begin{figure}[htb]
    \centering
    \includegraphics[width=0.7\textwidth]{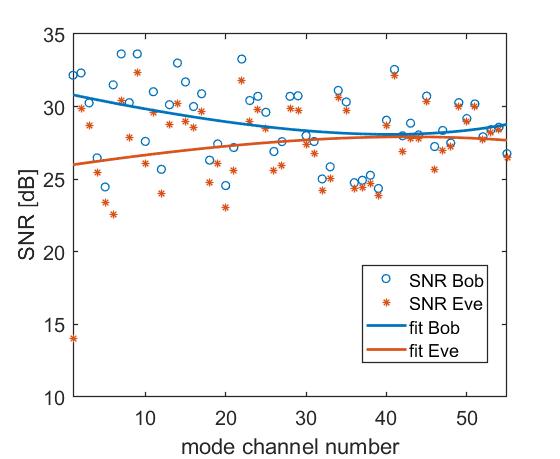}
    \caption{SNR plotted against the individual mode channels. Sequentially a digital '1' was transmitted over the mode channels and in Bob's and Eve's received signals the SNR was calculated, respectively.}
    \label{fig:SNR_no_noise}
\end{figure}

In the simulation the introduced communication, respectively eavesdropping model, is set up with the measured transmission matrices. In order to quantify the level of security using PLS, the quality of the two output signals $\vec{y}_B$ and $\vec{y}_E$ on Bob's and Eve's side is compared over the different mode channels with varying eavesdropping conditions. Both Bob and Eve perform thresholding during their detection, since dynamic channel changes are taken into account. While Bob can easily detect the message, due to Alice's inverse precoding [Eq.(\ref{eq:bob})], Eve compensates for the channel by a Tikhonov matrix inversion [Eq.(\ref{eq:outputeve})]. The \textit{Signal-To-Noise-Ratio} (SNR) of the detected signal is given as the performance parameter:
\begin{equation}
    \text{SNR}= 10 \cdot log_{10}\left(\frac{|\vec{y}_{i,\text{signal}}|^2}{|\vec{y}_{i,\text{background}}|^2}\right)
    \label{eq:SNR}
\end{equation}
First, we assume that the pure transmission characteristics are the same for both Bob and Eve. For this reason we assume for both the same measured $T_M$ from Figure \ref{fig:$T_M$}(a) ($T_{M,AE}=T_{M,AB}$). In addition, it is assumed for the simulation that the same additive white Gaussian noise occurs on both sides. If the MDL's were neglected, both output equations [Eq. (\ref{eq:bob}) and Eq. (\ref{eq:outputeve})] would be identical. However, in the presented communication model the MDL's are of course considered using the coupling matrix $V$. In \cite{winzer2015} two different coupling matrices were considered. It was assumed that the attenuation of modes is either i) logarithmic or ii) linear but always randomly distributed over the individual mode channels. 
In order to make an accurate selection of the Matrix $V$, power crosstalk was simulated using an \textit{Finite Difference Time Domain} (FDTD) based evanescent field coupling process \cite{Schmidt2013}. Individual modes of the MMF were excited in a straight MMF piece (case: transmission from Alice to Bob) and another straight MMF piece was placed next to it (case: coupling to Eve). This scenario was chosen based on the following consideration: Eve would splice the core of her MMF as close to the core of Alice's and Bob's MMF as she would receive sufficient intensity but would not be detected. The simulation results show that there is a deterministic relationship between the coupling behavior and the mode field distribution [see Supplement 1]. 
 The coupling process depends on the spatial distribution of the mode field power. Is the field power concentrated at the edge of the core, a particularly high proportion of the power is coupled to Eve's fiber ($6.5\%$ in the highest order mode), whereas particularly low power is coupled for lower order modes, where the main part of the power is guided in the center of the core ($0.018\%$ in the lowest order mode). For simplicity it is now assumed that during a splicing process the highest order mode is coupled without attenuation, i.e. a coupling factor of 1 is assigned to the last entry in the $V$ matrix. The lowest order mode now experiences an attenuation of $\frac{1}{6.5/0.018}=0.0028$, which is noted in the first entry of $V$. The attenuation values of the modes between these extreme values were now selected according to the following scheme: the proportion of the power of the field in the edge area in relation to the total total power of the core were calculated for all modes. The behaviour of this curve has now been scaled to the coupling factor values 0.0028 to 1. The result was taken as the diagonal entries of $V$ and can be seen in Figure \ref{fig:V}. The same $V$ matrix is assumed for all subsequent investigations.
If $V$ is now taken into account for the eavesdropping scenario, its influence can be investigated as follows. A digital '1' is sent successively over each of the 55 available mode channels and the output signals on both Bob's and Eve's side are compared with regard to the SNR [Eq. \ref{eq:SNR}]. In Figure \ref{fig:SNR_no_noise}, a significant difference can be observed on the mode channels on which a higher attenuation was present. On the channels where a lower or no attenuation was assumed, the SNR behaves almost the same for both Bob and Eve. Due to the fact that the $T_M$ of Eve was provided with attenuation weights, the entries of the $T_M$ at the locations with high attenuation are low in value. This causes a noise amplification during Eve's $T_M$ inversion performed during detection, whereby the values of the SNR decrease significantly of up to $\approx 20$dB in the worst case. This provides a crucial asymmetry between Eve and Bob.

\begin{figure}[htb]
\centering
(a)\includegraphics[width=0.6\textwidth]{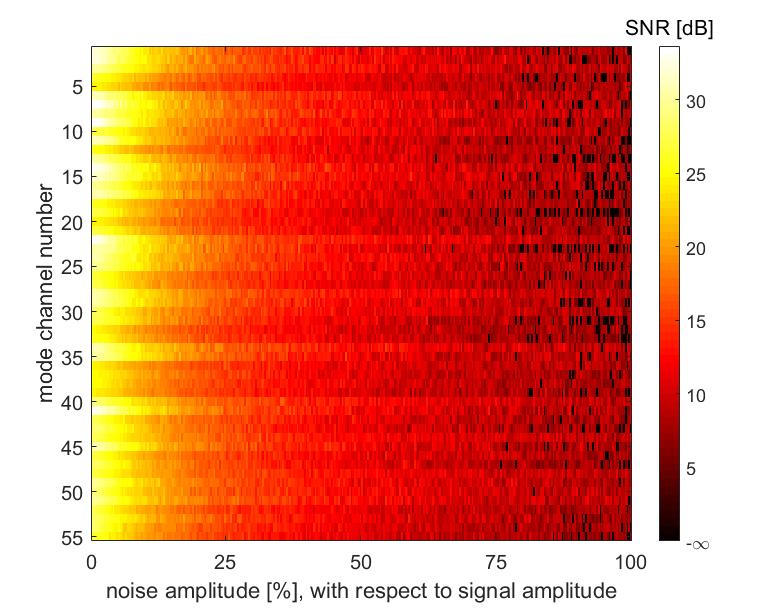} \\
(b)\includegraphics[width=0.6\textwidth]{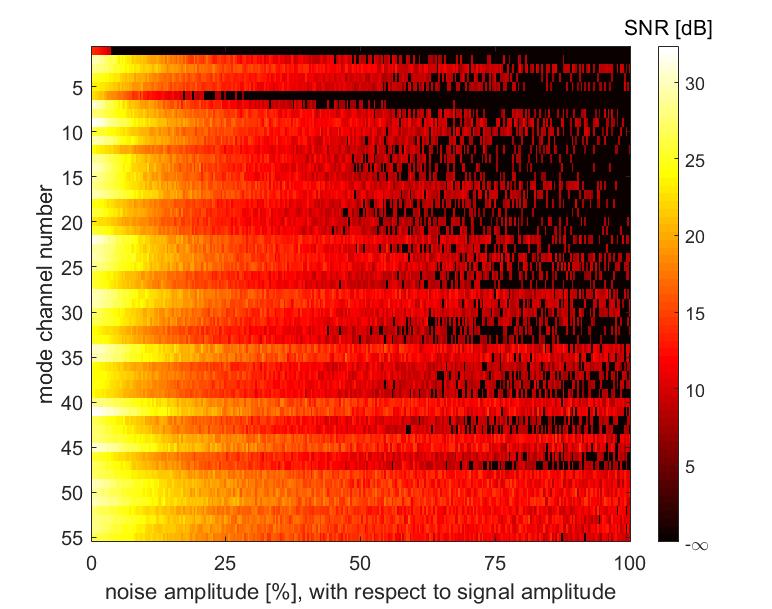}
\setlength{\belowcaptionskip}{0pt}
\caption{SNR for (a) Bob's and (b) Eve's detection respectively plotted over all 55 mode channels with increasing noise level. A '1' was sent sequentially over each channels with increasing noise level and both Bob's and Eve's  signals were evaluated. It should be noted that Eve multiplies her signal by the inverted $T_M$ $H^\dagger$ before evaluation, as shown in equation (\ref{eq:outputeve}). Bob, on the other hand, easily detects the signal sent by Alice due to her inverse precoding [Eq. (\ref{eq:bob})].}
\label{fig:SNR_noise}
\end{figure}

\begin{figure}[htb]
\centering
\includegraphics[width=0.6\textwidth]{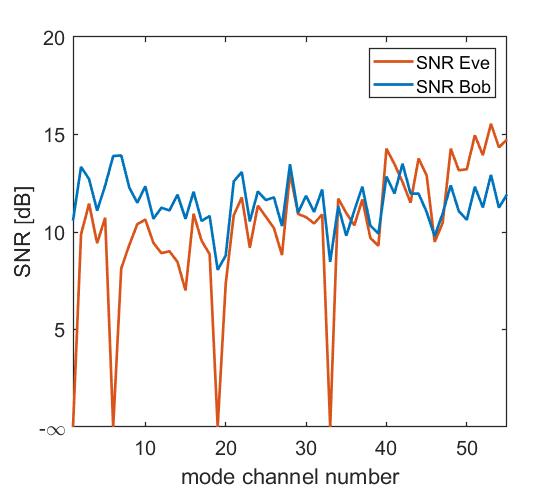}
\caption{A line scan from Figure \ref{fig:SNR_noise} at a noise level of $50\%$ at both Bob's and Eve's side. At this noise level, the mode channels 1, 6, 19, and 33 are safe, because Eve's detection on these channels is faulty. However, it can also be seen that Eve has a higher SNR than Bob for the signal detection of the mode channels that are particularly favourable for her (here, for example, channels 48 and up). This is due to the fact that Eve makes an inversion at the signal detection, which is not done by Bob. It is possible that the signal powers at the channels that are favourable for Eve are amplified and ultimately higher than on Bob's side.}
\label{fig:SNR_noise_50}
\end{figure}

 \noindent As mentioned in the previous section, the inverse precoding approach gives Alice the power to manipulate Eve's noise term. Alice thus can artificially add white Gaussian noise $\tilde{n}$ to the transmitted signal $\vec{x}_{A}$ from equation (\ref{eq:transmissionNew}):
 
\begin{equation}
    \tilde{\vec{x}}_{A}=T_{M,AB}^{\dagger}  (\vec{x}_{A}+\tilde{n}).
    \label{eq:transmissionNew2}
\end{equation}
The influence of the noise amplitudes on the mode channels can be deduced. Figure \ref{fig:SNR_noise} shows two plots where the colour-coded SNR was calculated for both Bob's and Eve's sides for every available mode channel with increasing noise amplitude (maximum $100\%$ with respect to signal amplitude), respectively. In the simulation, a digital '1' was again sent sequentially via all mode channels. As Bob and Eve use thresholding the SNR was only calculated for the case if the correct mode was detected. If the detection failed, the SNR value is set to $-\infty$ artificially. If the SNR drops to $-\infty$ on Eve's side, the channel is considered safe for Alice and Bob. As can be clearly seen in Figure \ref{fig:SNR_noise}, the amplitude of the artificial noise level has a significant influence on the signal quality at Eve's side. Bob, on the other hand, can almost consistently measure the correct signal, even if Alice adds a noise to her transmitted signal that has the same amplitude as the actual signal.
If one now performs a line scan in the two SNR evaluations shown in Figure \ref{fig:SNR_noise_50} at $50\%$ noise, it can be seen that Bob has an almost constant SNR level of 12dB. Additionally, there are 4 secure channels where Eve cannot decrypt the message sent by Alice, if Alice performs dynamic channel changes. With the presented technique the asymmetry between Bob and Eve is exploited to the highest degree, because Alice and Bob can communicate securely. By generating several secure channels it is also possible to increase the secure goodput, as it was introduced for wireless and optical communication channels in \cite{Lonnstrom2017}. One way to achieve this is by adjusting the artificial noise level, as the number of secure channels can be changed. 
\section{Secure information Transmission Using Mode division multiplexing}
\begin{figure}[htb]
\centering
(a)\includegraphics[width=0.6\textwidth]{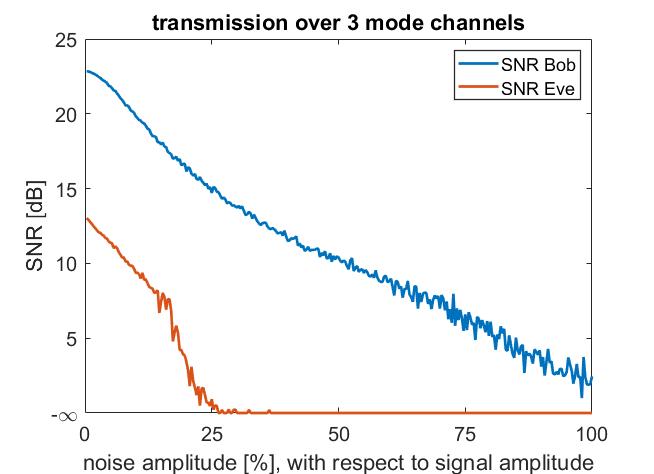}
(b)\includegraphics[width=0.6\textwidth]{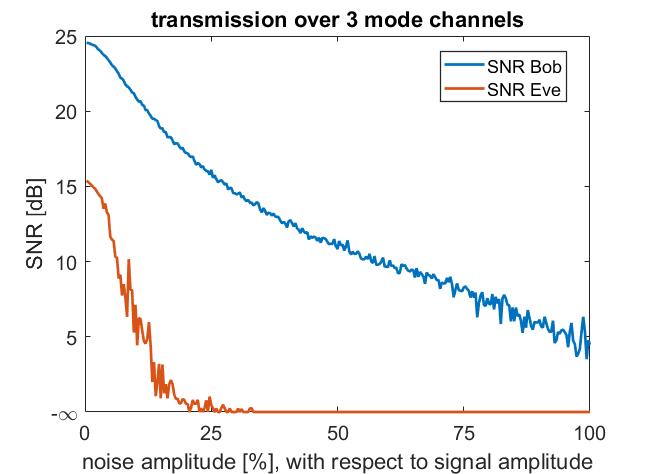}
\caption{SNR plotted against the increasing artificial noise level. Thresholding was used both on Bob's and Eve's side to detect the transmitted message (three times '1') (a) over three safe mode channels 1, 6 and 19 and (b) over two safe mode channels 1, 6, as well as one unsafe mode channel 50.}
\label{fig:SNR_MDM}
\end{figure}

In the previous section it was shown that at a noise level of $50\%$ there are 4 secure available channels for Alice to communicate over. In addition, it was shown that with the presented optical setup messages can be sent over several mode channels independently and simultaneously, or MDM can be performed via 3 modes. Now, that up to 3 modes can be transmitted simultaneously and safely [Fig. \ref{fig:SNR_MDM}(a)], it can be seen that on Eve's side, the detection at a noise level of about $50\%$ and above consistently fails, while Bob always receives the message in the correct way. It is also possible to transmit a message securely, even if not all individual channels are classified as secure [Fig. \ref{fig:SNR_MDM}(b)]. In the example shown here, it was always possible to transmit a message consisting of 3 bits securely via two secure channels and one arbitrary insecure channel. It is therefore possible to send messages with three channels from the entire mode domain as long as two channels, which are classified as safe, are involved in the message. Alice should add an artificial noise with a noise level of about $50\%$ of the signal amplitude to her signal to ensure security.
\section{Discussion}
The $T_M$ measurements shown here with subsequent inversion were performed on step-index MMFs of different lengths. The efficiency of diagonalization decreased from 95$\%$ at 10~cm to 89$\%$ at 2~m length. On the one hand, the decreasing efficiency can be explained by the fact that manufacturing tolerances play a greater role with increasing fiber lengths. The manufacturer specifies a tolerance of $10\%$ for both the NA and the core radius. On the other hand, MMFs with a step-index profile are said to be more difficult to control than, for example, MMFs with a graded-index profile \cite{Flaes2018}. For simplicity, a step-index profile was nevertheless used, since a mode domain of 55 modes per polarization at 532nm is to be expected here. The MMF with the smallest possible available core radius of $50\mu$m with a gradient-index profile has a mode domain with 1033 modes per polarization at 532nm. The goal of this paper was to proof the concept of PLS. For that reason, the focus so far has not been on using as many modes as possible, which is why using 1033 modes would go beyond the scope.

The mode-selective excitation ensures a $T_M$ having the smallest possible base. However, this technique is associated with a high adjustment effort, where smaller inaccuracies in the adjustment can induce a large extent of error. For example, the SLM is mapped to the input facet of the MMF at a scale of $\approx 1:600$. Other techniques using a Fourier pixels \cite{Loterie2015} or scanning focal points \cite{Ploschner2015} base are much easier to realize. However, in order to perform MDM, the measured $T_M$ would have to be translated back into the mode space via a conversion matrix, which increases the susceptibility to errors due to numerical inaccuracies. In addition, the number of measurements for a complete $T_M$ is much higher with alternative techniques, which is why the most elegant but also most difficult technique using mode-selective excitation was chosen for the setup presented here. 

The matrices measured with the introduced principle were used to perform an inverse precoding, so that communication can be done over arbitrary modes of the mode domain. Experimentally, it was shown that MDM can be performed with up to 3 independent modes via thresholding. Theoretically it would also be possible to communicate over more than 3 channels simultaneously, but this technique is limited by the inverse precoding quality. The number of SLM pixels used would be another limiting factor, but currently 130x130 superpixels are used for the excitation of modes, which should in principle suffice for the simultaneous excitation of all modes. 

The assumptions made for the eavesdropping scenario are favoring Eve's role. In reality, Eve is not almighty and the coupling coefficients are not scaled down to the very optimistic value 1, i.e. completely without attenuation. Even under this hypothesis, it could be shown that if Alice adds artificial noise with $50\%$ signal amplitude to her transmitted signal, there are 4 mode channels which can be classified as safe. The number of safe channels could, of course, change if the fiber parameters such as core radius or refractive index profile change.

\section{Conclusion}
Inverse precoding with artificial noise enables secure communication even if Eve knows the transmission matrices and is present during the calibration. 
The key for this is the exploitation of mode dependent losses which are characteristic for the channel behaviour inside MMFs. While Bob directly observes the sent modes, Eve needs to use matrix inversion. This gives Alice the power to introduce artificial noise and thus to amplify the noise Eve detects. This results in an SNR advantage for Bob, which is high enough to generate 4 mode channels in our exemplary simulation that can be considered to be secure as dynamic channel changes are used. In combination with MDM this is even enhanced. Messages can be sent safely over 3 channels from the entire mode domain, as long as 2 channels classified as safe are used. These results represent a crucial advancement for increasing the security in optical transmission channels with potential impact on data centers and cyber physical systems.
\section*{Funding Information}
This research was funded by Deutsche Forschungsgemeinschaft (DFG) grant numbers (CZ 55/42-1) and (JO 801/21-1).



\end{document}